\documentclass[reprint,amsmath,amssymb,aps,superscriptaddress, prl, longbibliography]{revtex4-1}
\usepackage{color}
\usepackage{graphicx}
\usepackage[colorlinks=true, urlcolor=blue, linkcolor=blue, citecolor=blue]{hyperref}
\usepackage{ulem}
\usepackage{footnote}

\usepackage{subfigure}
\usepackage[dvipsnames]{xcolor}

\begin{document}

\title{Observation of High Harmonics of the Cyclotron Resonance in Microwave Transmission of a High-Mobility Two-Dimensional Electron System}

\author{M.\,L.\,Savchenko}\email{mlsavchenko@isp.nsc.ru}
\affiliation{Rzhanov Institute of Semiconductor Physics, 630090 Novosibirsk, Russia}
\affiliation{Novosibirsk State University, 630090 Novosibirsk, Russia}

\author{A.\,Shuvaev}
\affiliation{Institute of Solid State Physics, Vienna University of Technology, 1040 Vienna, Austria}

\author{I.\,A.\,Dmitriev}
\affiliation{Terahertz Center, University of Regensburg, 93040 Regensburg, Germany}
\affiliation{Ioffe Institute, 194021 St.~Petersburg, Russia}

\author{A.\,A.\,Bykov}
\author{A.\,K.\,Bakarov}
\author{Z.\,D.\,Kvon}
\affiliation{Rzhanov Institute of Semiconductor Physics, 630090 Novosibirsk, Russia}
\affiliation{Novosibirsk State University, 630090 Novosibirsk, Russia}

\author{A.\,Pimenov}
\affiliation{Institute of Solid State Physics, Vienna University of Technology, 1040 Vienna, Austria}

\date{\today}

\begin{abstract}
We report an observation of magnetooscillations of the microwave power transmitted through the high mobility two-dimensional electron system hosted by a GaAs quantum well. 
The oscillations reflect an enhanced absorption of radiation at high harmonics of the cyclotron resonance and follow simultaneously measured microwave-induced resistance oscillations (MIRO) in the dc transport. 
While the relative amplitude (up to 1\%) of the transmittance oscillations appears to be small, they represent a significant ($>50$\%) modulation of the absorption coefficient. 
The analysis of obtained results demonstrates that the low-$B$ decay, magnitude, and polarization dependence of the transmittance oscillations accurately follow the theory describing photon-assisted scattering between distant disorder-broadened Landau levels. 
The extracted sample parameters reasonably well describe the concurrently measured MIRO. 
Our results provide an insight into the MIRO polarization immunity problem and pave the way to probe diverse high-frequency transport properties of high-mobility systems using precise transmission measurements. 
\end{abstract}

\maketitle

The discovery of the microwave-induced resistance oscillations (MIRO) and zero-resistance states (ZRS) in high-mobility two-dimensional electron systems (2DES) subject to a moderately strong perpendicular magnetic field~\cite{zudov:2001a,ye:2001,mani:2002,zudov:2003,dorozhkin:2003,yang:2003} triggered an outbreak of experimental and theoretical research that has led to observation of a number of interrelated magnetotransport phenomena in various materials and conditions~\cite{dmitriev:2012, zudov:2001b, yang:2002, zhang:2008, zhang:2007c, Bykov2008a, wiedmann:2008, wiedmann:2010b, willett:2004, smet:2005, bykov:2008d, dorozhkin:2009, dorozhkin:2011, Bykov2010a, andreev:2011, konstantinov:2010, konstantinov:2013, levin:2015, dorozhkin:2016, shi:2017, zudov:2014, karcher:2016, yamashiro:2015, zadorozhko:2018, otteneder:2018, friess:2020, tabrea:2020, moench:2020, kumaravadivel:2019}. Most of the observed effects have been coherently explained within a unified framework describing quantum kinetics of electrons in disorder-broadened Landau levels (LLs) in the presence of strong static and alternating electric fields~\cite{dmitriev:2012,ryzhii:1970,ryzhii:1986,durst:2003,dmitriev:2003,andreev:2003,vavilov:2004,ryzhii:2004,dmitriev:2005,vavilov:2007,khodas:2008,dmitriev:2009a,dmitriev:2009b,dmitriev:2010,raichev:2008,raichev:2009,monarkha:2019,dmitriev:2019,greenaway:2019,raichev:2020}
(for other theoretical proposals, see Refs.~[\onlinecite{dmitriev:2004,chepelianskii:2009,mikhailov:2011,zhirov:2013,beltukov:2016}]). 
In application to MIRO, this quantum description 
is tightly linked to an enhanced absorption of microwave radiation whenever the photon energy, $\hbar\omega$, is close to an integer multiple of the cyclotron energy, $N \hbar\omega_c$~\cite{dmitriev:2003,vavilov:2004,dmitriev:2005,raichev:2008}.
In a nutshell, such enhancements reflect the maxima of the thermally averaged product, $\langle\nu(\varepsilon)\nu(\varepsilon+\hbar\omega)\rangle$, of initial and final density of states $\nu(\varepsilon)$ for transitions between the disorder-broadened LLs. It is worth mentioning that, apart from the broadening of LLs, the role of disorder here is to make possible the otherwise dipole-forbidden photon-assisted transitions between distant LLs for $N\neq 1$.

The corresponding complex structure of the cyclotron resonance (CR) line shape  including multiple additional resonant features at $\omega=N\omega_c$, $N=2,3,\dots$, was observed already in the early far-infrared transmission experiments~\cite{abstreiter:1976}, following the theoretical predictions in pioneering works on quantum magnetotransport in 2DES~\cite{ando:1975a}.
However, a direct observation of the resonances associated with photon-assisted transitions between distant LLs turned out to be a challenging task in modern high-mobility 2DES~\cite{smet:2005,studenikin:2005,studenikin:2005a,wirthmann:2007,tung:2009,herrmann:2016}, despite giant MIRO and ZRS that are argued to emerge due to these processes
in the dc resistance of the microwave-illuminated samples~\cite{dmitriev:2003,vavilov:2004,dmitriev:2005}. 
As detailed below, the reason for this apparent contradiction is that the shape of the magneto-transmittance and reflectance in the ultra-high-mobility structures is dominated by the non-dissipative dynamical response of nearly free electrons, such that even very pronounced (up to $\sim 100\%$) quantum corrections to the dissipative part of the dynamic conductivity, $\mathrm{Re}\sigma(\omega, B)\propto\langle\nu(\varepsilon)\nu(\varepsilon+\hbar\omega)\rangle$, become hardly detectable in the transmission traces. 
It is thus not surprising that in high-mobility structures such quantum oscillations were so far only detected in the differential absorption of the 2DES placed in a cavity~\cite{fedorych:2010}. 

\begin{figure*}
	\includegraphics[width=2\columnwidth]{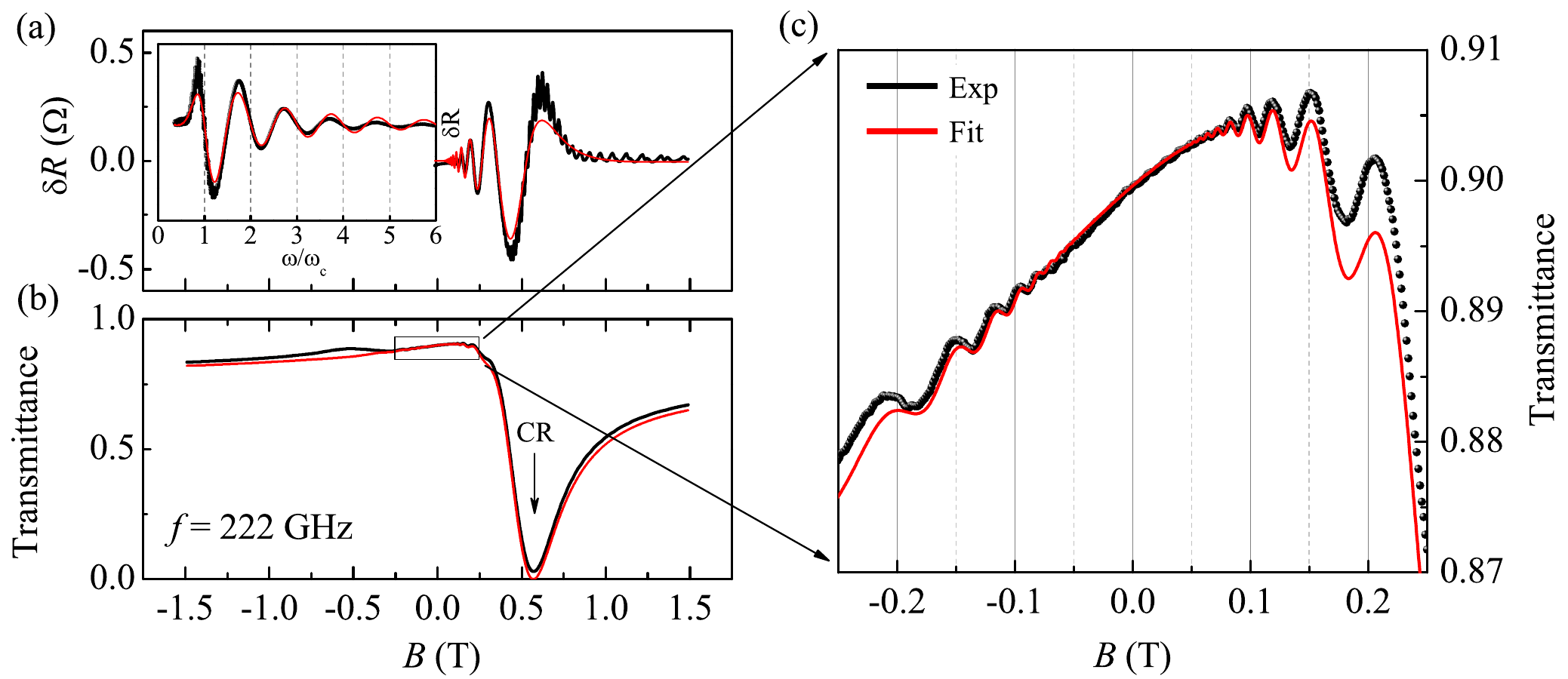}
	\caption{
		Magnetic field dependences of (a) photoresistance $\delta R(B)$ and (b) transmission $|t_+|^2(B)$ measured at frequency $f = 222\,$GHz and for the right-hand circular polarization (black).
		Inset in panel (a): a part of $\delta R(B)$ plotted against $\omega/\omega_\text{c}$.
		(c): a zoomed part of $|t_+|^2(B)$ in panel (b), where magnetooscillations in transmission are seen at both positive and negative magnetic fields. Red curves: fits calculated according to Eqs.~(\ref{Eq:T(B)}) -- (\ref{Eq:dSigma}) for the transmission and Eq.~(\ref{Eq:dR(B)}) for the photoresistance. 
	} \label{fig1}
\end{figure*}

Here 
we demonstrate the possibility to detect multiple harmonics of the CR in the transmission signal on a sample that also manifests well-pronounced MIRO in the in-situ measured dc resistance. To make this possible, we performed measurements on a sample with moderately high mobility, $\mu=2.1\times 10^6$\,cm$^2$/Vs, and implemented a quasi-optical setup with tunable-frequency stable radiation sources which enabled high accuracy measurements of the absolute transmittance values. To improve the visibility of the quantum magnetooscillations in transmittance 
we further took an opportunity to deliberately modify the shape of magneto-transmittance by fine tuning the interference effects originating from multiple reflections of the wave in the substrate. With all the above measures, we were able to resolve 8 or more periods of magnetooscillations for both polarities of the magnetic field under the circularly-polarized microwave illumination. We show that the low-$B$ decay, the magnitude, and polarization dependence of the observed oscillations accurately follow the theoretical predictions of Ref.~[\onlinecite{dmitriev:2003}], being governed by a single fitting parameter which describes the broadening of LLs due to impurity scattering. 
This parameter also reasonably well describes the low-$B$ decay of the concurrently measured MIRO. 
Our results demonstrate an opportunity to use high-precision transmission measurements to probe diverse high-frequency transport properties of high-mobility systems including quantum, hydrodynamic, or classical memory effects, and provide an additional insight into the intriguing and controversial issue of polarization immunity of MIRO~\cite{smet:2005,herrmann:2016,zadorozhko:2018,dmitriev:2012,chepelianskii:2018}.

\noindent
{\it Sample and methods.}-- The measurements were carried out on a heterostructure containing a 2DES in a selectively doped 16\,nm GaAs quantum well with AlAs/GaAs superlattice barriers grown by molecular beam epitaxy on a GaAs substrate \cite{baba:1983,Friedland1996,umansky:2009,manfra:2014}. The van-der-Pauw sample size was 10$\times$10\,mm, ohmic contacts at the corners were fabricated by burning Ge/Au/Ni/Au, see Supplemental Material (SM) for details~\cite{SM1}. 
After exposure to the room light the electron density and mobility were $n = 6.6\times10^{11}\,$cm$^{-2}$ and  $\mu = 2.1\times 10^{6}\,$cm$^{2}$/Vs, respectively~\cite{SM1}. The sample was irradiated from the substrate side through an 8\,mm aperture. Backward-wave oscillators with available frequency range between 100 and $1000$\,GHz were used as stable sources of the normally incident continuous monochromatic radiation. A split-coil superconducting magnet provided the magnetic field, $B$, oriented perpendicular to the sample surface. The transmittance through the sample was measured using a He-cooled bolometer. High temporal stability of the whole system including the mounting of the sample allows us to detect the relative changes of transmittance down to $10^{-4}$~\cite{shuvaev_sst_2012, SM1}. In parallel with the transmittance, the photoresistance $\delta R$ (the difference of the resistance signals in the presence and absence of irradiation) was measured using the double-modulation technique~\cite{SM1}. All presented results were obtained at temperature $T=1.9$\,K.

\noindent{\it Results.} -- Figure \ref{fig1} shows a representative example of simultaneously measured photoresistance, panel (a), and transmittance, panels (b) and (c), recorded as a function of perpendicular magnetic field $B$ using $f=222$\,GHz circularly  polarized radiation. Red lines in Fig.~\ref{fig1} are theoretical fits, explained below, to the data shown by black lines and dots. Similar results obtained under 328 and 432\,GHz radiation are provided in SM~\cite{SM1}.

The photoresistance $\delta R$ in panel (a) displays pronounced MIRO governed by the ratio $\omega/\omega_\text{c}$, where
$\omega = 2 \pi f$ is the angular radiation frequency and $\omega_\text{c} = e B/m$ is the cyclotron frequency. Here $e=|e|$ is the
elementary charge and $m$ is the electron effective mass. In the inset, a part of these data is replotted versus $\omega/\omega_\text{c}$ (using $m = 0.068\,m_0$, where $m_0$ is a free electron mass) making evident that the observed magnetooscillations  
accurately reproduce the established period and phase of MIRO, with nodes at integer $\omega/\omega_\text{c}$. 

The focus of this work is on the magnetotransmittance.
The analysis below shows that it is dominated by strong metallic reflection from the 2DES leading to a strong dip in the region of CR, marked by the arrow in panel~(b). The transmission is also significantly affected by multiple reflections of the electromagnetic wave in the dielectric slab between back and front interfaces of the sample. The associated Fabry-P\'{e}rot interference is responsible for an asymmetric line shape of the CR dip in transmittance. For our particular choice of microwave frequency, $f=222$\,GHz, the interference produces an almost flat shoulder at $|B|\lesssim 0.2$\,T. A magnified view of this region, shown in panel (c), reveals the presence of magnetooscillations which closely resemble the $\omega/\omega_\text{c}$-oscillations seen in $\delta R$. Despite the relatively low amplitude of magnetooscillations in transmittance and their exponential decay towards low $B$, we are able to clearly detect multiple oscillation periods for both polarities of the magnetic field. Below we demonstrate that these oscillations are in excellent agreement with theoretical predictions of Ref.~[\onlinecite{dmitriev:2003}], see red line in panel (c), and can therefore by attributed to an enhanced absorption at integer $\omega/\omega_\text{c}$ due to resonant photon-assisted transitions between distant disorder-broadened LLs.

\noindent{\it Analysis.} -- We model the measured transmittance using the expression~\cite{abstreiter:1976,herrmann:2016}
\begin{equation}\label{Eq:T(B)}
|t_\pm|^2 = \dfrac{4 K_\text{sw}}{\left|s_1 \left(1 + \sigma_\pm Z_0\right) + s_2\right|^{2}}\,.
\end{equation}
Apart from a phenomenological factor $K_\text{sw}\simeq 1$ addressed below, it describes the fraction of power transmitted through a dielectric slab containing an isotropic 2DES for a normally incident circularly-polarized wave ($+$ and $-$ signs correspond to the right- and left-handed circular polarization, respectively). 
Two complex parameters $s_1=\cos(k d)-i\epsilon^{-1/2}\sin(k d)$ and $s_2=\cos(k d)-i\sqrt{\epsilon}\sin(k d)$ describe the Fabry-Per\'ot interference and are controlled by the product of the sample thickness, $d=406\,\mu$m, and the wave number in the GaAs substrate, $k=\sqrt{\epsilon}\omega/c$. 
These parameters were accurately determined from the period of measured frequency dependence of transmittance at $B=0$, yielding the dielectric permittivity $\epsilon=12.06$~\cite{SM1}. 
Apart from that, these $B=0$ measurements provided the value of the $\omega$-dependent factor $K_\text{sw}$ (0.913 for the fit in Fig.~\ref{fig1}) which is conventionally introduced to account for weak uncontrolled effects such as remaining standing waves in the experimental cell~\cite{SM1,dziom_phd_2018}.

The magnetic field dependence of $|t_\pm|^2$ in Eq.~(\ref{Eq:T(B)}) is fully determined by the complex dynamic conductivity of the 2DES, $\sigma_\pm$, which enters in combination with the impedance of free space $Z_0$ and is defined as $\sigma_\pm=\sigma_{xx}(\omega, B)\pm i\sigma_{xy}(\omega, B)$ in terms of components $\sigma_{xx}=\sigma_{yy}$ and $\sigma_{xy}=-\sigma_{yx}$ of the conductivity tensor. We have checked that all features of the measured $|t_\pm|^2$ remain independent of the microwave power in the whole available range, which demonstrates that our measurements reflect the linear-response transport properties of the 2DES~\cite{SM1}. 
The linear conductivity is modelled~\cite{dmitriev:2003,SM1} as a sum $\sigma_\pm=\sigma_\pm^\text{D}+\delta\sigma_\pm$ of classical Drude conductivity,
\begin{equation}\label{Eq:Sigma(B)}
\sigma^\text{D}_\pm = \frac{e n}{\mu^{-1} \pm i B- i B_\text{CR} },
\end{equation}
where $B_\text{CR}=\omega m_\text{CR}/e$ denotes the position of the CR,
and an oscillatory quantum correction,
\begin{equation}\label{Eq:dSigma}
    \delta\sigma_\pm = 
    2\delta^2\cos\frac{2\pi \omega}{\omega_\text{c}}\,\text{Re}\,\sigma^\text{D}_\pm.
\end{equation}
Here $\delta = \exp \left( -\pi /\mu_\text{q} |B| \right)$ is the Dingle factor, and $\mu_\text{q}$ is the quantum mobility which parameterizes the disorder broadening of LLs. For a reliable comparison of the theoretical model with experiment, in the transmittance fit according to Eqs.~(\ref{Eq:T(B)})-(\ref{Eq:dSigma}), presented in Fig.~\ref{fig1}, we use the electron mobility $\mu = 2.1\times 10^{6}\,$cm$^{2}$/Vs and density $n = 6.6\times10^{11}\,$cm$^{-2}$ determined from the dc magnetotransport measurements~\cite{SM1}. 

The remaining three fitting parameters $B_\text{CR}$, $m=eB/\omega_\text{c}$, and $\mu_\text{q}$ can be independently extracted from the transmission data. Namely, the CR position $B_\text{CR}$ can be accurately determined from the position of deep minimum in panel (b) of Fig.~(\ref{fig1}), where $|t_+|^2$ approaches zero in view of a large value of $e n \mu Z_0\simeq 84$ in our high-mobility 2DES. The corresponding cyclotron mass, $m_\text{CR}=0.073\,m_0$
represents the collective cyclotron motion of electrons which by virtue of the Kohn theorem~\cite{kohn:1961} is insensitive to electron-electron interactions. In contrast, the quasiparticle effective mass, $m=0.068\,m_0$, determined from the period of $\omega/\omega_\text{c}$-magnetooscillations in both the microwave transmission and dc resistance, represents the LL spacing in the vicinity of the Fermi level where the microwave-assisted scattering processes take place. This mass should therefore be sensitive to the Fermi-liquid renormalizations and can significantly differ from $m_\text{CR}$ as confirmed by previous MIRO observations in several materials~\cite{hatke:2013,shchepetilnikov:2017,fu:2017,tabrea:2020}. Our data involve a large number of oscillation periods in both $\delta R$ and $|t_+|^2$ and thus enable a precise determination of $m$. The results presented in Fig.~\ref{fig1} demonstrate that the periods of MIRO and of quantum oscillations (Eq.~(\ref{Eq:dSigma})) in the dynamic conductivity coincide. Taking into account that, unlike MIRO, quantum oscillations in transmission become more pronounced for lower transport mobility, this opens up an opportunity to explore the renormalization effects~\cite{hatke:2013,shchepetilnikov:2017,fu:2017,tabrea:2020} in a broader class of 2DES.  

The form of Eq.~(\ref{Eq:dSigma}) suggests that knowledge of a single parameter, $\mu_\text{q}$, should be sufficient to reproduce not only the low-$B$ decay, but also the shape and magnitude of magnetooscillations in the relative quantum correction $\delta\sigma_\pm/\text{Re}\,\sigma^\text{D}_\pm$ to the dissipative part of dynamic Drude conductivity. Taking into account that the values of all other relevant parameters were fixed using the results of independent measurements~\cite{SM1}, this makes the accuracy of the fit in panel (c) of Fig.~(\ref{fig1}) quite remarkable. The obtained value of $\mu_\text{q}=0.23 \times 10^6$\,cm$^2$/Vs is approximately 10 times smaller than the transport mobility, which implies the dominance of small-angle impurity scattering typical for modern high-mobility 2DES. The validity of the above description is additionally supported by fits of the transmission data obtained for higher frequencies that yield similar values of $\mu_\text{q}$~\cite{SM1}.

The same value of $\mu_\text{q}$ reasonably well describes the low-$B$ decay of MIRO in $\delta R$, which are modelled using the theoretical expression~\cite{dmitriev:2009b} 
\begin{equation}\label{Eq:dR(B)}
\frac{\delta R}{R} = -A_\omega \frac{ \omega}{\omega_\text{c}} \delta^2 \sin \frac{2\pi \omega}{\omega_\text{c}},
\end{equation}
see red line in panel (a). The factor $A_\omega$ here depends on several parameters including the radiation characteristics~\cite{herrmann:2016} and properties of 2DES~\cite{SM1}. 
The estimates show that MIRO at relevant temperatures should be dominated by the inelastic mechanism~\cite{dmitriev:2003, dmitriev:2005}. 
The fit provided the required value of the microwave power of 0.2\,mW, which is consistent with the characteristics of implemented radiation sources~\cite{SM1}. 
A closer look at high harmonics of MIRO shows, however, that the value of $\mu_\text{q}$ found from the transmittance oscillations underestimates the actual decay of MIRO. 
We found that MIRO can be well fitted using a shorter value of $\mu_\text{q}=0.135 \times 10^6$\,cm$^2$/Vs which, in turn, does not fit the magnitude and number of oscillations in the transmittance~\cite{SM1}. 
This unexpected discrepancy requires understanding and deserves additional studies.  

\noindent{\it Discussion.} -- While the relative amplitude $\sim 1\%$ of the observed oscillations in $|t_+|^2$ is quite small, they represent a significant effect ($2\delta^2\sim 0.5$ at $|B|=0.2$\,T) in the dissipative part of the dynamic conductivity, $\text{Re}\,\sigma_{+}$, which defines the fraction of the microwave power absorbed by 2DES. The reason is that outside an extremely narrow range $|B-B_\text{CR}|\lesssim \mu^{-1}\sim 0.05$\,T near the CR,
the shape of the magnetotransmission is controlled by the imaginary part of conductivity,
 $\mathrm{Im} \sigma_{+} \simeq e n/(B_\text{CR}-B)$, 
which describes classical forced oscillations of electrons in the microwave field in the absence of scattering and thus is insensitive to Landau quantization~\cite{SM1}. 
The real part, proportional to the rate of photon-assisted scattering off impurities, is modified as the thermally averaged product, $\langle\nu(\varepsilon)\nu(\varepsilon+\hbar\omega)\rangle$, of initial and final density of states $\nu(\varepsilon)$ for transitions between the disorder-broadened LLs. For the relevant high $T$, corresponding to suppressed Shubnikov-de Haas oscillations, 
and in the limit of strongly overlapping LLs, $\nu(\varepsilon)/\nu_0\simeq 1 - 2\delta \cos(2\pi\varepsilon/\hbar\omega_\text{c})$, where $\nu_0$ denotes the constant density of states at $B=0$, this average reduces to $\nu_0^2 \left(1+2\delta^2\cos(2\pi\omega/\omega_\text{c}) \right)$, reproducing Eq.~(\ref{Eq:dSigma}) \cite{SM1,dmitriev:2003,dmitriev:2012}. 

The approximation of overlapping LLs works well for $\delta\ll 1$, but underestimates the amplitude of magnetooscillations in Fig.~\ref{fig1} for $|B|\gtrsim0.2$\,T, corresponding to transition to the regime of separated LLs. More importantly, the shape of magnetotransmission at $|B|\gtrsim0.2$\,T is also influenced by such uncontrolled effects as remaining standing waves in the experimental cell~\cite{SM1}. These detrimental effects are, apparently, the main cause of deviations of the fitting curve from measured $|t_+|^2$ both at $B\sim B_\text{CR}$ and at $B\sim -B_\text{CR}$, which makes difficult a reliable comparison between the obtained data and theory in these interesting ranges of $B$. Importantly, in the well controlled range of $|B|<0.2$\,T illustrated in panel (c), we do not observe any  deviations between the theoretical fit and data that are asymmetric with respect to the sign of $B$. This suggests that the polarization dependence of the measured transmission $|t_+|^2$, and, therefore, of the dynamic conductivity is well captured by theory~\cite{dmitriev:2003}, as opposed to MIRO, where strong deviations were reported for GaAs-based 2DES~\cite{smet:2005,herrmann:2016}, while a recent experiment on electron system on the surface of liquid He demonstrated the polarization dependence consistent with the theory predictions~\cite{zadorozhko:2018}. 

Similar to the Shubnikov-de Haas oscillations in the dc transport response, quantum oscillations in transmission discussed above are a direct manifestation of the Landau quantization, and thus should emerge in any 2DES in the appropriate range of radiation frequencies and magnetic fields. Classical memory effects can be another potential source of strong $\omega/\omega_\text{c}$-oscillations in high-frequency dissipative transport~\cite{polyakov:2002,dmitriev:2004,beltukov:2016,dorozhkin:2017} that can  be studied using high precision measurements of transmission. Unlike universal quantum oscillations, the shape, phase, and damping of these classical oscillations are strongly sensitive to the type of random potential of impurities realized in a particular 2DES. One further interesting potential application concerns hydrodynamic and magnetoplasmon effects in high-frequency transport, which, in particular, are proposed~\cite{alekseev:2019,volkov:2014} to be responsible for a huge photoresistance peak observed at $\omega\simeq2\omega_\text{c}$ \cite{dai:2010,hatke:2011b}.

In conclusion, we have observed resonant transmission features at multiple integer harmonics of the cyclotron resonance which follow the microwave induced resistance oscillations concurrently measured in dc magnetotransport. The detected transmission oscillations, including their polarization dependence, accurately follow  the theoretical predictions relating them to impurity- and photon-assisted transitions between distant Landau levels, thus supporting the 
quantum mechanisms proposed for explanation of MIRO. Our work demonstrates that high precision measurements of transmission can be a versatile tool for studies of sufficiently strong effects in high-frequency dissipative transport of high-mobility 2DES.

\begin{acknowledgments}
We acknowledge discussions with D.A.~Kozlov and support from Austrian Science Funds I3456-N27, from the Russian Foundation for Basic Research (17-52-14007), and from the German Research Foundation (DFG project GA501/14-1).
\end{acknowledgments}

\bibliography{newbib}
\end{document}


\title{Supplemental Material to:\\ 
		``Observation of High Harmonics of the Cyclotron Resonance in Microwave Transmission of a High-Mobility Two-Dimensional Electron System''}
	
\author{M.\,L.\,Savchenko}
\affiliation{Rzhanov Institute of Semiconductor Physics, 630090 Novosibirsk, Russia}
\affiliation{Novosibirsk State University, 630090 Novosibirsk, Russia}

\author{A.\,Shuvaev}
\affiliation{Institute of Solid State Physics, Vienna University of Technology, 1040 Vienna, Austria}

\author{I.\,A.\,Dmitriev}
\affiliation{Terahertz Center, University of Regensburg, 93040 Regensburg, Germany}
\affiliation{Ioffe Institute, 194021 St.~Petersburg, Russia}

\author{A.\,A.\,Bykov}
\author{A.\,K.\,Bakarov}
\author{Z.\,D.\,Kvon}
\affiliation{Rzhanov Institute of Semiconductor Physics, 630090 Novosibirsk, Russia}
\affiliation{Novosibirsk State University, 630090 Novosibirsk, Russia}

\author{A.\,Pimenov}
\affiliation{Institute of Solid State Physics, Vienna University of Technology, 1040 Vienna, Austria}
	
	\date{\today}
	\begin{abstract}
	    In this supplemental material we provide details of transmittance measurements, additional transmission data for different radiation frequencies, dependence of MIRO and transmittance on the radiation power and dc current, magnetotransport in the absence and presence of radiation, MIRO measurements using double modulation technique, full expression for the MIRO amplitude, summary of the fitting procedure including table of obtained sample parameters, discussion of the low-$B$ decay of oscillations in transmission and MIRO, and conditions for observation of transmittance oscillations in high-mobility 2DES.
	\end{abstract}
	
	\maketitle
	
\setcounter{figure}{0}
\renewcommand{\thesection}{S\arabic{section}}
\renewcommand{\theequation} {S\arabic{equation}}
\renewcommand{\thefigure} {S\arabic{figure}}
\renewcommand{\thetable} {S\arabic{table}}
	
		\section{Details of transmittance measurements}
	\label{sec_tr}
	
	The sample insert consists of a fixed part with the beam aperture and the movable part with the sample. The sample is fixed on a thin (6 to 13\,$\mu$m) Mylar foil which is clipped to the movable rod of the insert. This is done to have as few movable elements around the sample as possible, in order to avoid unwanted disturbances of the beam. The detected signal as a function of radiation frequency is measured twice: first time without the sample, yielding the reference spectrum	$I_\text{ref}(f)$, and second time in the presence of the sample, which gives $I_\text{sam}(f)$. The power transmission coefficient through the sample is then obtained as $T(f) = |t_\pm|^2(f) = I_\text{sam}(f) / I_\text{ref}(f)$. Here, $t_\pm$ is the complex transmission coefficient (see Eq.~(1) of the main text), and $+(-)$ denotes the right (left) hand circular polarization (note that $|t_+|^2=|t_-|^2$ at $B=0$).
	
\begin{figure}[b]
		\includegraphics[width=0.8\columnwidth]{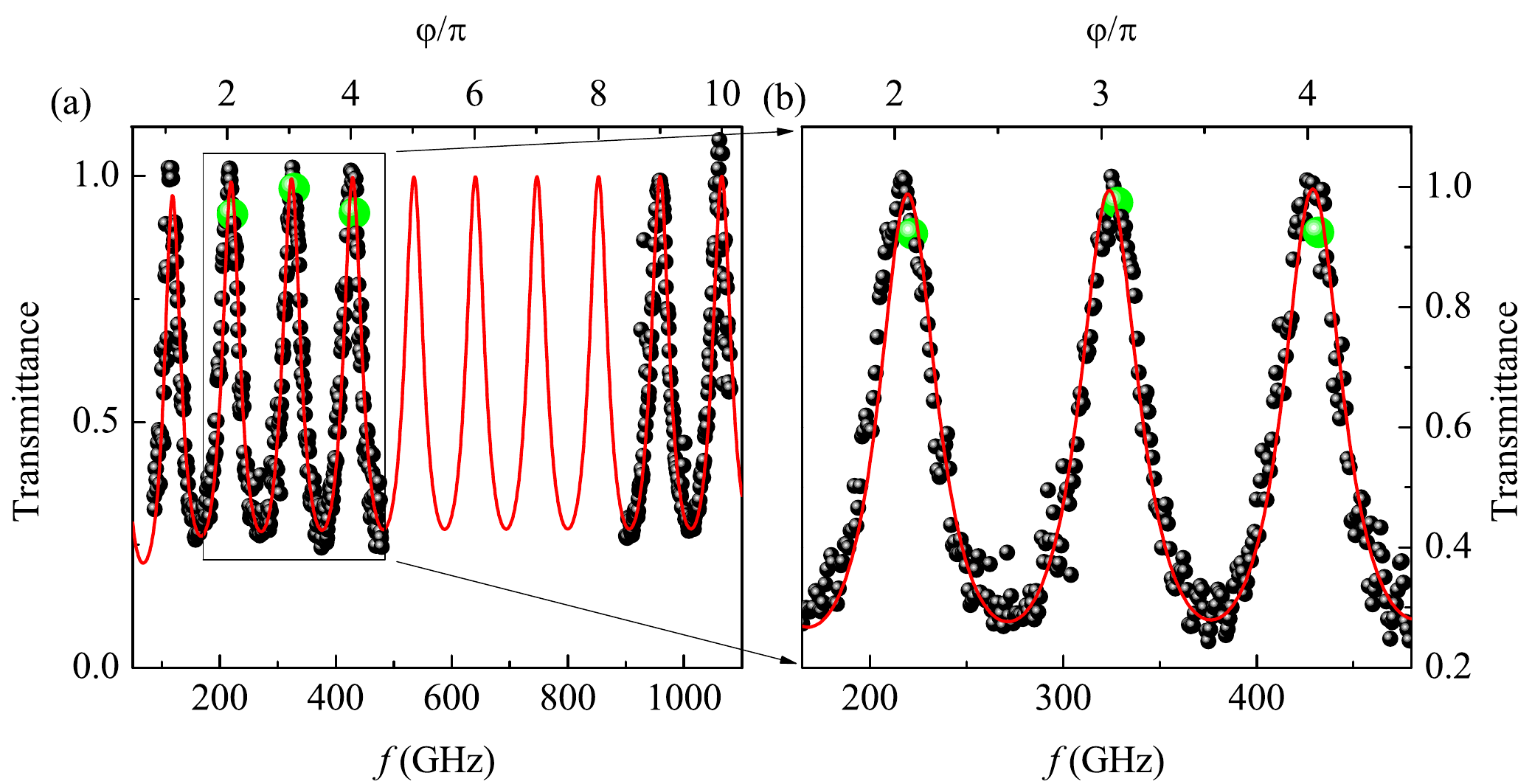}
		\caption{
			Transmittance $T(f)$ at $B=0$ as a function of radiation frequency $f$ (bottom axis).
			The top axis shows the corresponding Fabry-Per\'ot interference phase $\varphi\equiv k d= 2 \pi  f d\sqrt{\epsilon}/c$ in units of $\pi$. Symbols represent the experimental data points, while the red line shows a fit according to Eqs.~(1) -- (2) of the main text. 
			Big green spheres correspond to frequencies at which presented $|t|^2(B)$ curves are measured. (b): A zoomed region of panel (a) for the relevant interval of frequencies.
		} \label{FigS1}
	\end{figure}

	In Fig.~\ref{FigS1} we show the frequency dependence of transmittance $T(f)$ at $B=0$ which clearly demonstrates the Fabry-P\'{e}rot interference due to internal reflections between back and front interfaces of the sample (see Eq.~(1) of the main text).	Black spheres represent the experimental data, while the red line is calculated according to Eqs. (1) and (2) of the main text (e.g., using the Drude model (2) for the dynamic conductivity). Big green spheres correspond to three frequencies chosen for measurements presented in this work. These frequencies have been chosen for two reasons. First, they correspond to the most stable ranges of the radiation sources. Second, the analysis according to Eqs.~(1) -- (3) of the main text shows that transmittance is most sensitive to the relevant conductivity corrections in the magnetic field region, where the transmittance itself reaches maximum value close to unity (this property is further illustrated in Sec. \ref{Conditions}).	Since the $\omega/\omega_\text{c}$-periodic transmittance oscillations are expected to occur at low magnetic fields, the chosen frequencies are blue shifted a little bit from the values corresponding to complete transparency of the dielectric substrate  (i.e., from the maxima in $T(f)$ occurring due to the constructive Fabry-P\'{e}rot interference at integer $\varphi/\pi$).	In this case the combination of the interference constants $s_1$ and $s_2$, characterizing the dielectric substrate at a given $f$, and additional reflection from the 2DES results in transmittance values close to unity in the desired range of low magnetic fields.
	
	The theoretical description of transmission in Eq.~(1) of the main text assumes that the portions of the electromagnetic wave transmitted through or reflected from the sample never return back. In this case, the unity ratio between the detector signal in the absence and presence of the sample indeed corresponds to the unity transmission through the sample. In reality, however, a part of radiation is reflected back from various optical elements and still reaches the detector. The frequency-dependent interference of these secondary beams with each other and with the primary beam leads to the formation of a complex structure of maxima and minima in the observed spectrum and to visible deviations from the expected interference behavior described by Eq.~(1). These irregular oscillations are superimposed to the intrinsic spectrum of the sample giving it a ``noisy'' look, see Fig.~\ref{FigS1}.	
	It should be mentioned, however, that the observed complex transmission pattern is highly reproducible and can easily be obtained, e.g., on next day,	provided the measurement arm has not been modified in between. The limited sample volume in the Oxford cryomagnet (26\,mm bore) with a lot of metallic cladding increases the effects of the standing waves. 
	In case of the frequency dependent spectra the spurious standing waves
	cancel each other in average as they change the transmission both in positive and negative directions. However, for single frequency measurements, especially when high precision is required as in the present case, the influence of these standing waves can be considerable. In order to account for them, the additional factor $K_\text{sw} \simeq 1$	is introduced in the Eq.~(1) of the main text. This factor, however, neglects the variations of the $B$-dependence of transmission introduced by secondary waves passing through the 2DES with $B$-dependent conductivity. Despite all precautions made, currently this limits our ability for a more precise analysis of the transmission data obtained near $|B|\sim B_\text{CR}$, probably also affecting the data near the second harmonics of the CR. For these reasons, the analysis and comparison to the theoretical predictions in the main text is limited to the region of  $|B|\lesssim 0.2$ where the spurious effects of the secondary waves (or, more precisely, of our inaccurate account for them using the constant factor $K_\text{sw}$) are expected to be negligible.
	
	\newpage
	\section{Transmission oscillations at different radiation frequencies}
	\begin{figure}[h]
		\includegraphics[width=0.9\columnwidth]{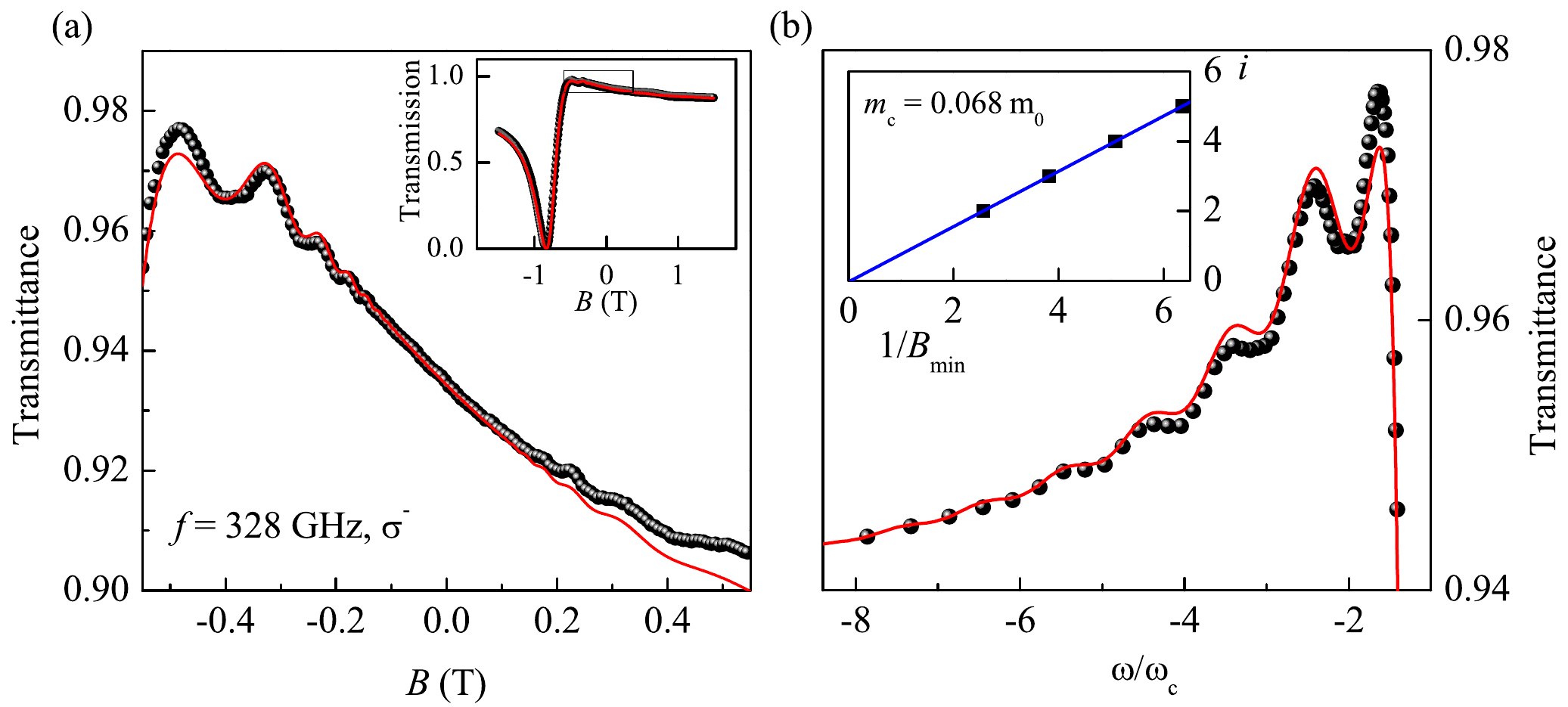}
		\caption{
			(a) A zoomed part of the magnetic field dependence of transmittance $|t_-|^2$ measured at frequency $f = 328\,$GHz and left-hand circular polarization (black). The transmittance oscillations are seen at both negative and positive magnetic fields. The red curve is calculated according to Eqs.~(1) -- (4) of the main text with parameters given in Table \ref{Table}. Inset: transmittance $|t_-|^2$ in the full range of magnetic field.	(b) The $B<0$ portion of the data in left panel plotted against $\omega/\omega_\text{c}$. Inset: reciprocal magnetic
			fields of the $|t_-|^2$ oscillations minima plotted against their number.
		}\label{FigS2}
	\end{figure}
	
	\begin{figure}[h]
		\includegraphics[width=0.85\columnwidth]{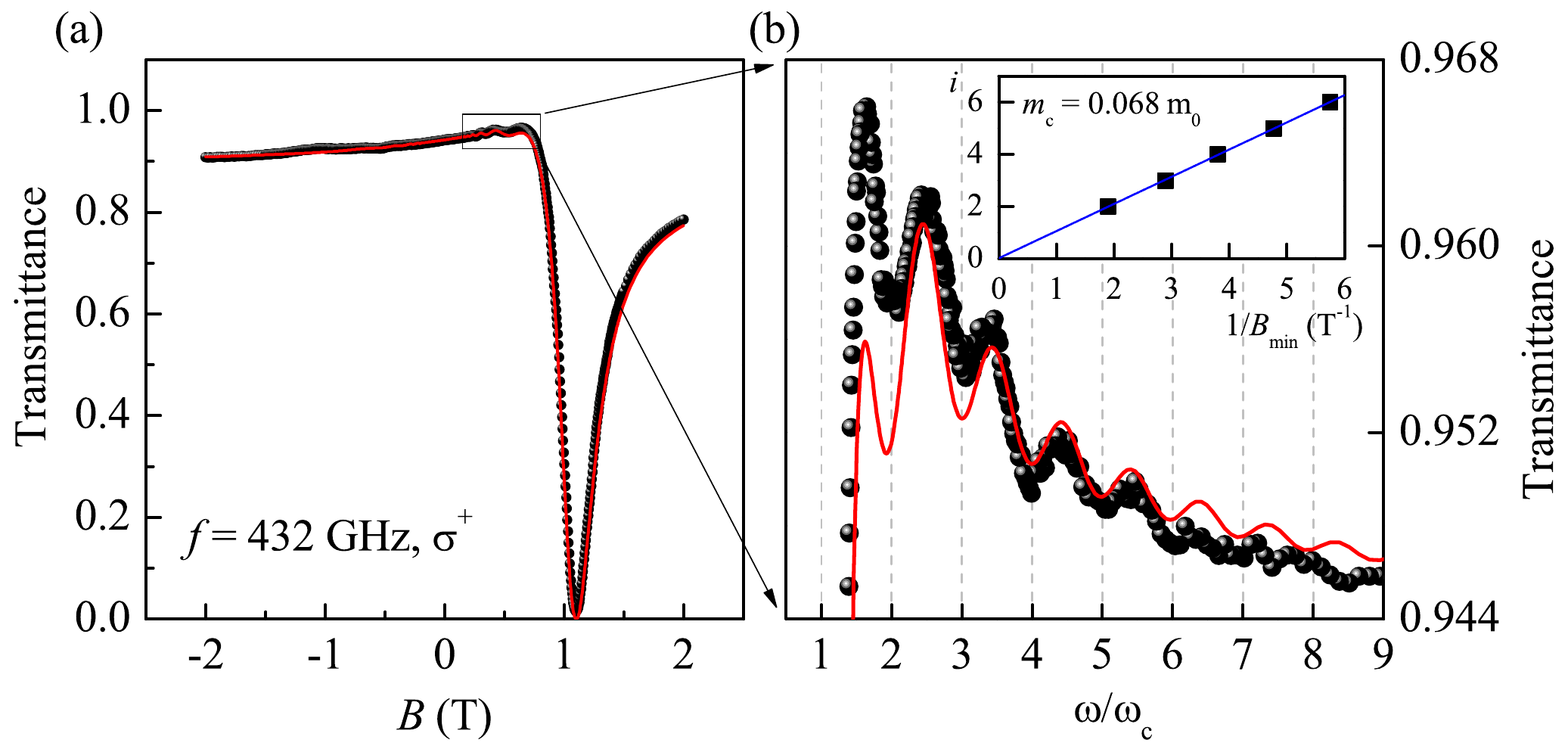}
		\caption{
			(a) The magnetic field dependence of transmittance $|t_+|^2$ measured at the frequency $f = 432\,$GHz and right-hand circular polarization~(black). Both the cyclotron resonance and magnetooscillations are seen. The red curve is calculated according to Eqs.~(1) -- (4) of the main text with parameters given in Table \ref{Table}.
			(b) A zoomed part of the dependence $|t_+|^2$ plotted against $\omega/\omega_\text{c}$. Inset: reciprocal magnetic
			fields of the $|t_-|^2$ oscillations minima plotted against their number.
		} \label{FigS3}
	\end{figure}
	
	\newpage
	\section{Data for different power and dc current}
	\label{power}
	\begin{figure}[h]
		\includegraphics[width=0.9\columnwidth]{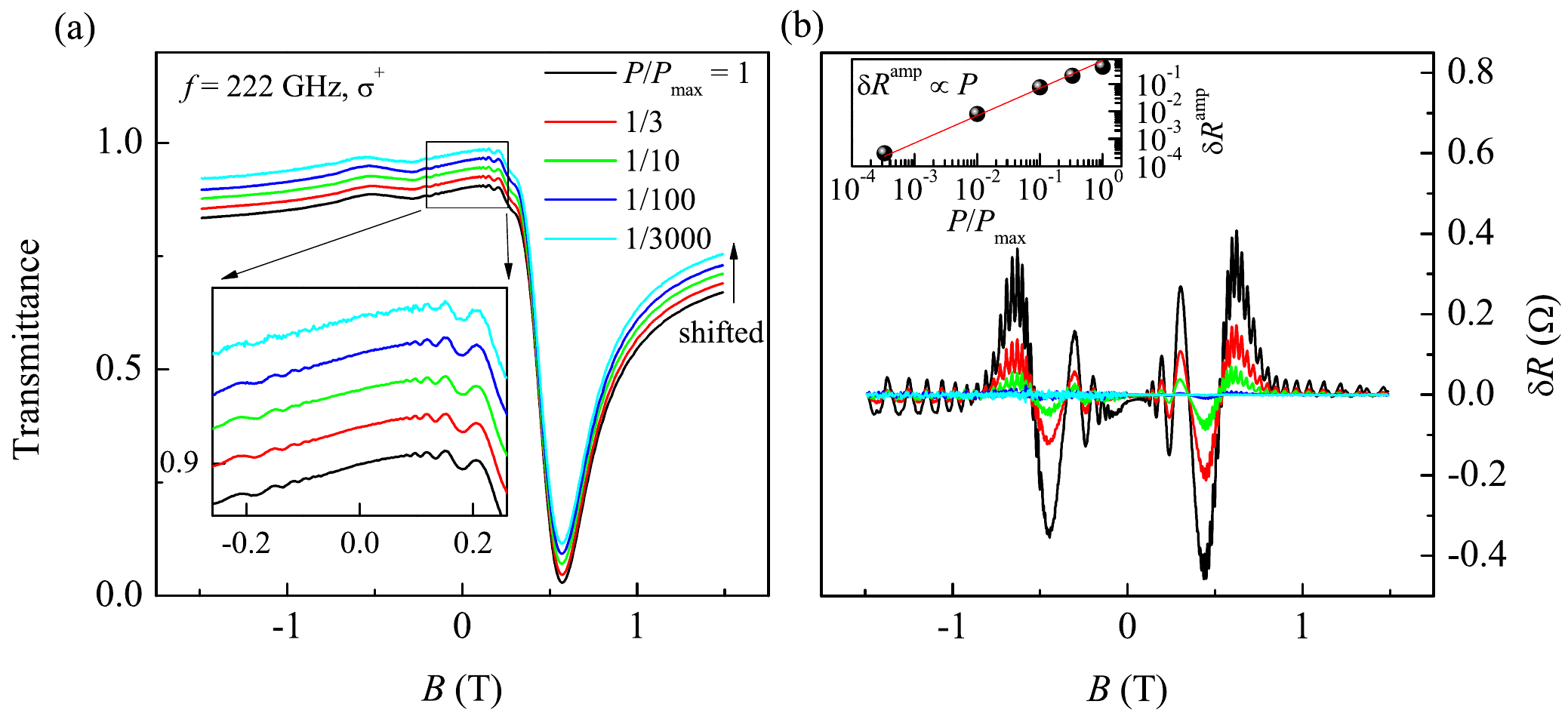}
		\caption{
			Magnetic field dependences of transmittance $|t_+|^2(B)$~(a) and photoresistance $\delta R(B)$~(b) measured at frequency $f = 222\,$GHz and right-hand circular polarization for different levels of the radiation power $P$ (in units of full power $P_\text{max}$, see legend).
			The transmittance curves are vertically shifted for clarity. The transmittance does not show any nonlinear effects: The whole dependence including the region of magnetooscillations [zoomed in panel (a)] remains the same for all power levels. As demonstrated in inset in panel (b), the amplitude of MIRO $\delta R^\text{amp}$ is proportional to $P$, with a sublinear deviation barely noticeable only at the highest available power $P_\text{max}$.   
		} \label{FigS4}
	\end{figure}
	
	\begin{figure}[h]
		\includegraphics[width=0.8\columnwidth]{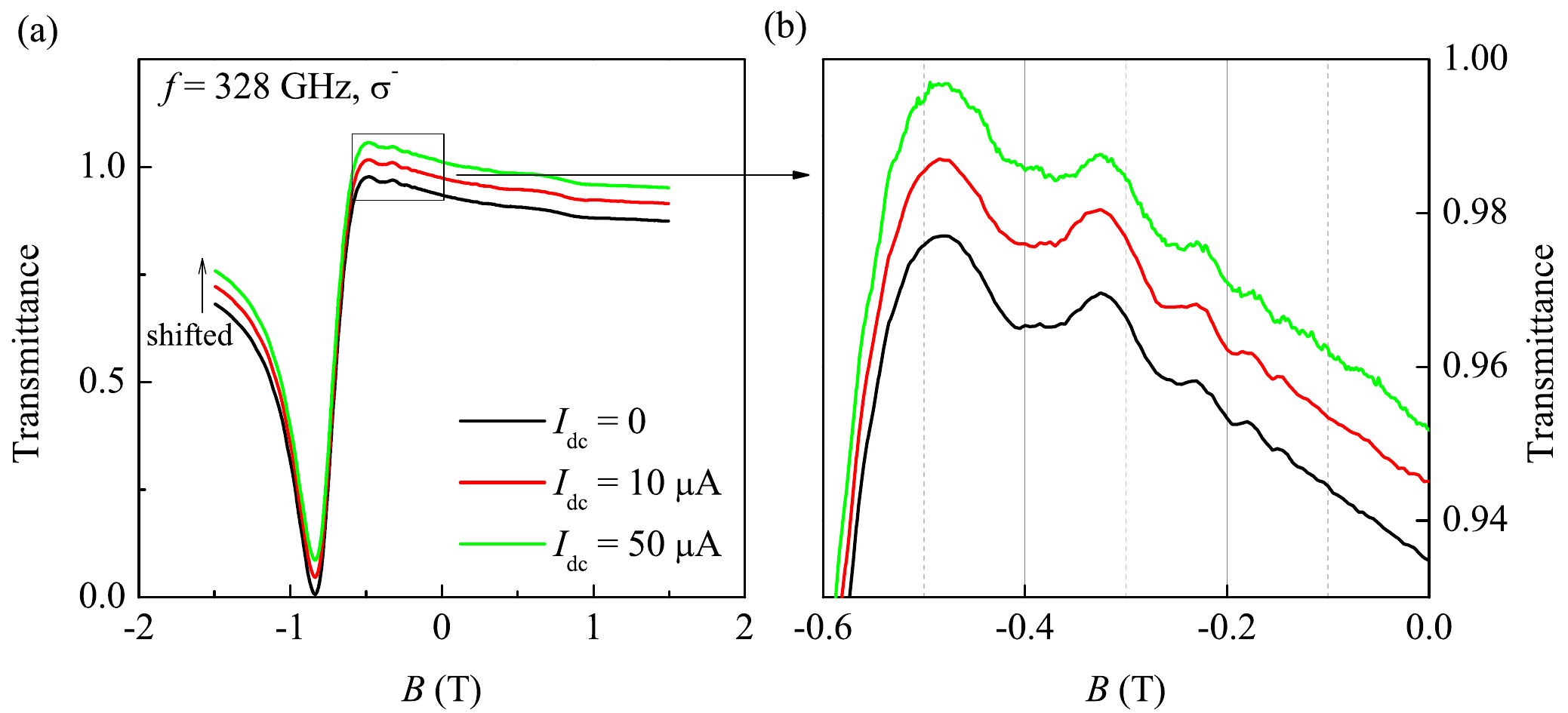}
		\caption{
			Magnetic field dependences of transmittance $|t_-|^2$ measured at frequency $f = 328\,$GHz and left-hand circular polarization at different dc bias currents $I_\text{dc} = 0,~10,~50\,\mu$A. The transmittance curves are vertically shifted for clarity. As expected, the whole dependence including the region of magnetooscillations [zoomed in panel (b)] remains the same under application of small dc bias used for concurrent transport and MIRO measurements.
		} \label{FigS5}
	\end{figure}

	\newpage
	\section{Magnetotransport in the absence and presence of radiation}
	\label{sTransport}

\begin{figure}[h]
		\includegraphics[width=1\columnwidth]{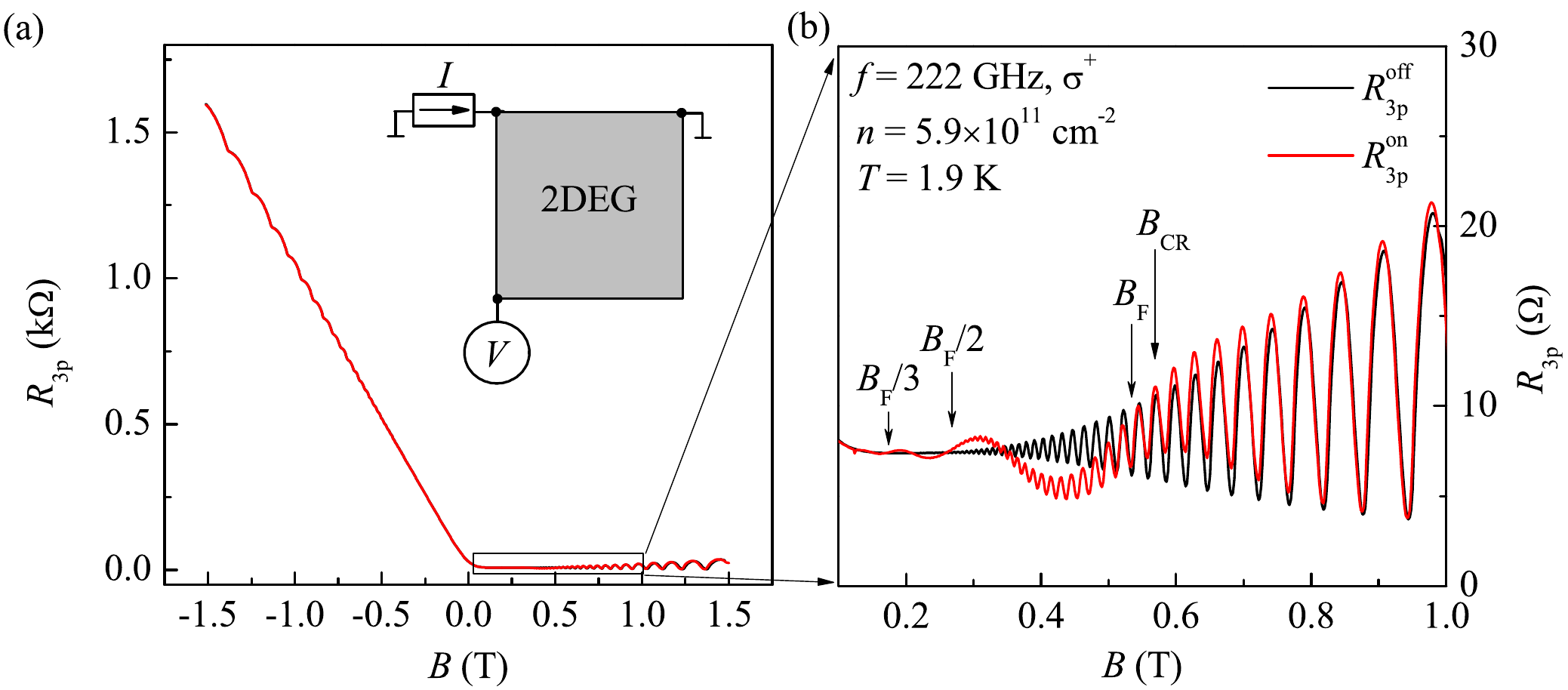}
		\caption{
			(a)~Magnetic field dependence of measured three-point resistance $R_\text{3p}$ with ($R_\text{3p}^\text{on}$, red line) and without ($R_\text{3p}^\text{off}$, black line) illumination with a right-hand circularly polarized $f = 222\,$GHz microwave radiation. (b) MIRO in a zoomed part of the dependence in left panel. An arrow marked $B_\text{CR}=2\pi f m_\text{CR}/e$ indicates the position of the cyclotron resonance obtained from the fits of measured transmittance using classical Drude formula. The nodes of MIRO appear at $B=B_\text{F}/N$ with $N=1,2,$\,\ldots, corresponding to $\delta R(B)\propto -\sin(2\pi B_\text{F}/B)$, where the fundamental frequency $B_\text{F}=2\pi f m/e$ includes the quasiparticle effective mass $m$ different from the cyclotron mass $m_\text{CR}$ (see Sec.~\ref{Sec: Fitting procedure}).} 
			\label{FigS6}
	\end{figure}

	Magnetotransport measurements were performed using a standard low-frequency lock-in technique with a driving current $I_\text{dc} = 1 - 50\,\mu$A at a frequency of 12\,Hz. 
	The data presented in the paper were obtained on the sample in a Van der Pauw	geometry where only three corner contacts were available, see a schematic image in Fig.~\ref{FigS6}\,(a).
	Figure \ref{FigS6} shows the three-point resistance $R_\text{3p}$ measured in the absence (black line) and presence (red line) of the $f = 222\,$GHz radiation. MIRO are clearly seen in the zoomed part of $R_\text{3p}(B)$ dependence at positive $B$ in Fig.~\ref{FigS6}\,(b) as an additional long period modulation of the resistivity between 0.2 and 0.6\,T in the presence of illumination. 
	The short period oscillations present both with and without radiation are Shubnikov -- de Haas oscillations.
	It is evident that $R_\text{3p}$ is strongly asymmetric, being dominated by the diagonal resistivity $\rho_{xx}$ of 2DEG at $B>0$ and by the Hall resistivity $\rho_{xy}$ at $B<0$.	 
	The unknown relative contributions of $\rho_{xx}$ and	$\rho_{xy}$ to the signal at $B<0$ make it impossible to reliably analyze the corresponding part of the photoresistance $\delta R(B)$ obtained using the double modulation technique. Therefore, the analysis of MIRO is limited to the region of $B>0$.
	The slope of $R_\text{3p}$ at negative $B$ gives the electron density consistent with a more accurate value (see Table \ref{Table}) obtained from the period of Shubnikov -- de Haas oscillations. 
	To obtain the mobility values listed in the Table we used a Hall-bar structure made from the same wafer. 
	
	\newpage
	\section{MIRO measurements using double modulation technique}
	
	Usually the microwave-induced resistance oscillations (MIRO) are measured under continuous microwave illumination. 
	The photoresponce is thereby obtained by comparing the magnetic field dependencies $R^\text{on}$ and $R^\text{off}$ of the longitudinal resistance in the presence and absence of radiation. 
	But at low radiation power the amplitude of MIRO becomes small, and this approach is not optimal. 
	Moreover, our main goal here is to probe the transmittance of radiation that is measured using mechanical chopper such that the incoming radiation is modulated at frequency $f_\text{chopper}$. 
	Therefore, we measure photoresistance $\delta R = R^\text{on} - R^\text{off}$ using a double modulation technique specified below.
	
\begin{figure}[h]
		\includegraphics[width=0.7\columnwidth]{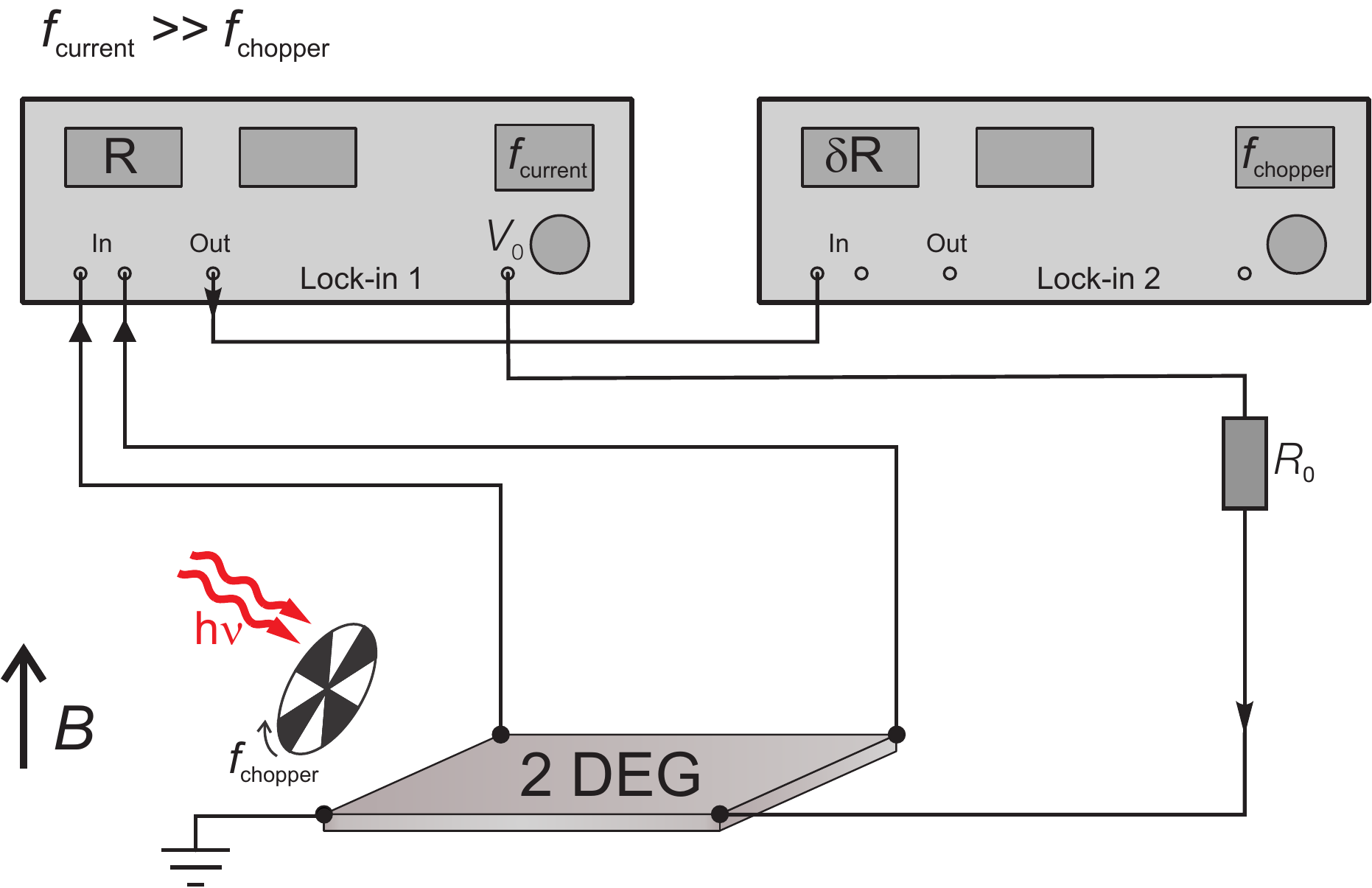}
		\caption{
			Illustration of the double-modulation technique used in the measurements of photoresistance $\delta R$.
			Here the chopper frequency $f_\text{chopper} =23\,$Hz and the bias current is applied at higher frequency
			$f_\text{current} = 1.1\,$kHz. The difference photoresistance signal $\delta R$ is obtained by using two lock-ins in series, where the first one collects the signal which oscillates in phase with the applied bias at frequency $f_\text{current} = 1.1\,$kHz, while the second collects the part $\delta R = R^\text{on} - R^\text{off}$ of the resulting dc resistance which oscillates in phase with the microwave power modulated at frequency $f_\text{chopper} =23\,$Hz.
		} \label{FigS7}
	\end{figure}

	In our measurements, see Fig.~\ref{FigS7}, the chopper frequency $f_\text{chopper} =23\,$Hz and a small bias current $I_\text{dc} = 1 - 50\,\mu$A is applied to the sample at much higher frequency $f_\text{current} = 1.1\,$kHz.
	The photoresistance is collected as a part of signal that is modulated at both frequencies: The first lock-in collects the total signal at higher frequency $f_\text{current}$, while its 	output is fed to the input of the second lock-in tuned to the lower frequency $f_\text{chopper}$. The output signal of the second lock-in
	gives the value of $\delta R$. 
	In such measurement scheme, such amplifier settings as integration time and a band filter slope have to be properly adjusted to the modulation frequencies $f_\text{current}$ and $f_\text{chopper}$.
	
	\newpage
	\section{MIRO amplitude}
	\label{MIRO}
	
	We use the following expression to fit MIRO (for more details, see Supplemental materials of~\cite{herrmann:2016}),
	\begin{equation}\label{Eq:dR(B)_full}
	\frac{\delta R}{R} = -\eta P S_\pm \frac{2\pi \omega}{\omega_\text{c}} \delta^2 \sin \frac{2\pi \omega}{\omega_\text{c}},
	\end{equation}
	where 
	\begin{equation*}
		P = \frac{4 \pi e^2 n I}{c\, \epsilon_0 m^2 \omega^4},\qquad  S_\pm = \frac{  |t_\pm|^2 }{1/(\omega \tau)^2 + (1 \mp \omega_\text{c}/\omega)^2}.
	\end{equation*}
	Here $I=P_\text{inc}/S_\text{ill}$ is the radiation intensity which is estimated as the incident radiation power, $P_\text{inc}$, divided by the illuminated area of the sample, $S_\text{ill}=50$~mm$^2$, given by the area of the aperture. The factor $\eta$ includes contributions of the displacement and inelastic mechanisms of MIRO, and is expected to be dominated by the inelastic contribution if $\eta_\text{in} \approx 2 \pi \hbar^3 n/m^* \tau (k_\text{B}T)^2$ takes values much larger than unity \cite{dmitriev:2009b}.	In our case calculation gives $\eta_\text{in} \simeq 20$, so one may conclude that MIRO is governed by the inelastic mechanism,
	$\eta \approx \eta_\text{in}$. 
	In the main text, we combine the three factors entering Eq.~(\ref{Eq:dR(B)_full}) as $A_\omega = 2 \pi \eta P S_\pm$. Using Eq.~(\ref{Eq:dR(B)_full}) for fitting MIRO, we obtain the only unknown parameter -- the incident power $P_\text{inc}$. The resulting values of $P_\text{inc}$, given in Table~\ref{Table}, are consistent with the nominal power of our radiation sources.

	\newpage
	\section{Summary of fitting procedure and table of sample parameters}
	\label{Sec: Fitting procedure}
		
	Below we review the procedure used to determine all possible sample parameters in equations (1)-(4) of the main text from independent measurements, thus reducing the number of free fitting parameters. The remaining free parameters can be independently and reliably determined from distinct features of the measured magnetotransmittance. The values of obtained parameters are summarized in Table~\ref{Table}. 
	
	\begin{enumerate}
	\item The Fabry-P\'{e}rot interference parameters $s_1$ and $s_2$ from Eq.~(1) of the main text require knowledge of the sample thickness $d = 406\,\mu$m, measured by a micrometer, and of the dielectric constant of the substrate, $\epsilon=12.06$, which is obtained from the period of the Fabry-P\'{e}rot interference in the frequency dependence of transmittance measured at zero magnetic field, see Fig.~\ref{FigS1}. The same dependence gives the factor $K_\text{sw}$ for a given measurement frequency (see Sec.~\ref{sec_tr} for details).
	
	\item The electron density $n$ and mobility $\mu$ of the 2DES entering the Drude formula, Eq.~(2) of the main text, are determined from the dc magnetotransport measurements (see Sec.~\ref{sTransport} for details).
	
	\item Taking into account the sign-alternating character and small amplitude of observed magnetooscillations in transmittance, the measured dependences $|t|^2(B)$ for different radiation frequencies $f$ are first fitted using Eqs.~(1) and (2), i.e., neglecting the oscillatory correction introduced in Eq.~(3). 
	With other parameters entering Eqs.~(1) and (2) being fixed as described above, the only remaining fitting parameter at this step is the cyclotron mass, $m_\text{CR}$, which defines the position $B_\text{CR}=2\pi f m_\text{CR}/e$ of the CR in Eq.~(2).

	\item Reciprocal magnetic fields that correspond to the	extrema of transmittance oscillations  (see, e.g., insets of Fig.~\ref{FigS2}\,(b) and \ref{FigS3}\,(b) for 328 and 432\,GHz, respectively) are plotted against their sequential number. The slope of the obtained linear dependence gives the value of the effective electron mass $m$ entering $\omega_c=e B/m$ in Eqs.~(3) and (4). As expected, the obtained value of $m$ is different from $m_\text{CR}$, which can be attributed to the Fermi-liquid renormalization by electron-electron interactions [see comparison of $B_\text{CR}=2\pi f m_\text{CR}/e$ and $B_\text{F}=2\pi f m/e$ in Fig.~\ref{FigS6}\,(b)]. 
	It is important to mention that we observe a large number of oscillation periods both in the transmittance and in dc resistance which enables an accurate determination of the period and phase of oscillations. The analysis shows that the observed oscillations accurately reproduce Eqs.~(3) and (4) which predict $|t|^2\propto \cos(2\pi B_\text{F}/B)$ and $\delta R\propto -\sin(2\pi B_\text{F}/B)$, with no noticeable deviations between the values of $m$ extracted from the period of MIRO and from the transmittance oscillations.  
		
	\item 
	The amplitude and low-$B$ decay of the transmittance oscillations are fitted using the remaining free parameter, the quantum mobility $\mu_\text{q}$. As expected, the obtained value is about one order of magnitude lower than the transport mobility, and also reasonably well describes the low-$B$ decay of MIRO. A more careful analysis shows, however, that the decay of MIRO can be better described with a somewhat lower value of $\mu_\text{q}$, see Sec.~\ref{alt_fitting}.
		
	\item Everywhere apart from Sec.~\ref{alt_fitting}, the parameters determined above are also used for fitting MIRO. Here the only remaining free parameter is the incident radiation power, $P_\text{inc}$, which is used to fit the amplitude of MIRO, see also Secs.~\ref{MIRO} and \ref{power}. The resulting values of $P_\text{inc}$, see Table~\ref{Table}, are consistent with the nominal power of our radiation sources. 
		
	\end{enumerate}

	\begin{table}[h]
		\caption{\label{Table}%
			Parameters of the sample determined from the experiment. Fixed parameters
			were the sample thickness $d = 406\,\mu$m, temperature $T=1.9$\,K, and the illuminated area 
			$S_\text{ill}=50$\,mm$^2$. }
		\begin{ruledtabular}
			\begin{tabular}{cccccccc}
				$f$ & $\epsilon$& $m_\text{CR}$& $m$& $n$& $\mu$& $\mu_\text{q}$& $P_\text{inc}$ \\
				\hline
				222\,GHz\footnote{
				First line represents the data presented in the main text, in Fig.~\ref{FigS4}, and in Fig.~\ref{FigS8}, in which cases the sample was illuminated by room light before measurements. All other data were obtained without such illumination.
				}& 12.06&  0.0725 m$_0$& 0.0675 m$_0$& 6.6$\times 10^{11}$\,cm$^{-2}$& 2.1$\times 10^6$\,cm$^{2}$/Vs&
				0.23$\times 10^6$\,cm$^{2}$/Vs& 0.2\,mW\\
				\hline
				222\,GHz& 12.06& 0.072 m$_0$& 0.068 m$_0$& 5.9$\times 10^{11}$\,cm$^{-2}$& 1.8$\times 10^6$\,cm$^{2}$/Vs&
				0.17$\times 10^6$\,cm$^{2}$/Vs& 0.2\,mW \\
				\hline
				328\,GHz& 12.06& 0.072 m$_0$& 0.068 m$_0$& 5.9$\times 10^{11}$\,cm$^{-2}$& 1.8$\times 10^6$\,cm$^{2}$/Vs&
				0.15$\times 10^6$\,cm$^{2}$/Vs& 0.02\,mW \\
				\hline
				432\,GHz& 12.06& 0.072 m$_0$& 0.068 m$_0$& 5.9$\times 10^{11}$\,cm$^{-2}$& 1.8$\times 10^6$\,cm$^{2}$/Vs&
				0.2$\times 10^6$\,cm$^{2}$/Vs& -\\
			\end{tabular}
		\end{ruledtabular}
	\end{table}

	\newpage
	\section{Low-$B$ decay of transmittance oscillations and MIRO}
	\label{alt_fitting}

\begin{figure}[h]
		\includegraphics[width=0.7\columnwidth]{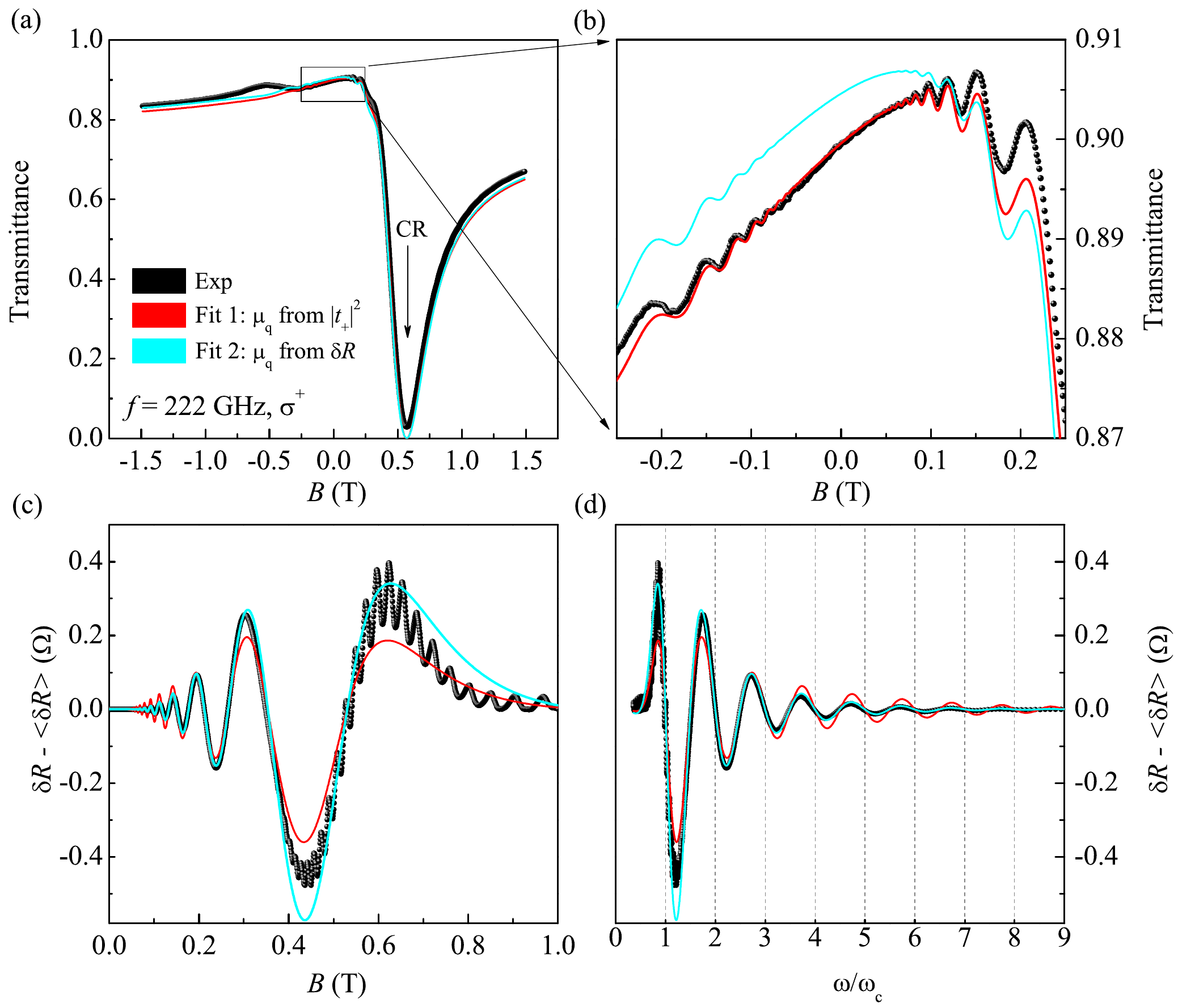}
		\caption{
		Black symbols: Magnetic field dependences of transmittance $|t_+|^2$ [panels (a) and (b)] and of the photoresistance $\delta R - \langle\delta R\rangle$ [with subtracted smooth background $\langle\delta R\rangle$, panels (c) and (d)] measured at frequency $f = 222\,$GHz and for right-hand polarization.	Red and cyan lines: Fits according to Eq.~(1) -- (3) of the main text for the transmittance and to Eq.~(4) for the photoresistance. Experimental data and red fits are the same as in Fig.~1 of the main text. The red fits are calculated using 
		$\mu = 2.1\times 10^6$\,cm$^{2}$/Vs and $\mu_\text{q} = 0.23\times 10^6$\,cm$^{2}$/Vs, while the cyan fits -- using $\mu = 0.7\times 10^6$\,cm$^{2}$/Vs and $\mu_\text{q} = 0.135\times 10^6$\,cm$^{2}$/Vs.
		} \label{FigS8}
	\end{figure}

	The procedure described in Sec.~\ref{Sec: Fitting procedure} allowed us to determine all sample parameters one by one in a well-defined sequential order, thus avoiding unreliable multi-parameter fitting. In final steps of this procedure, we first determined the quantum mobility $\mu_\text{q}$ from the transmittance oscillations, and then used the resulting value for modelling MIRO (where the only remaining fitting parameter was the common constant factor, proportional to the microwave power). Comparison of the experimental traces for transmittance and MIRO (black lines in Fig.~\ref{FigS8} and in Fig.~1 of the main text) and the resulting theoretical fits (red lines) with $\mu_\text{q} = 0.23\times 10^6$\,cm$^{2}$/Vs demonstrates a precise agreement for the transmittance, but also reveals a noticeable underestimation of the experimental MIRO damping, most evident for high harmonics of MIRO in Fig.~\ref{FigS8}\,(d). 
	
	Here we present the results of an alternative analysis (cyan lines) where $\mu_\text{q}= 0.135\times 10^6$\,cm$^{2}$/Vs is rather determined from the low-$B$ decay of MIRO as it is conventionally done when no results for the transmittance oscillations are available. With such lower $\mu_\text{q}$, MIRO can be perfectly fitted in the whole range of $B$ including higher harmonics, but the modelled transmittance oscillations both decay much faster than in experiment and become much smaller in magnitude (by a factor of three at $B\sim 0.2$~T).  
	In order to make oscillations in transmittance visible in Fig.~\ref{FigS8}\,(b), for the cyan fit we arbitrarily used a three times smaller value $\mu = 0.7\times 10^6$\,cm$^{2}$/Vs of the transport mobility, which makes the magnitude of oscillations in transmission similar to that observed in experiment. The comparison of the red and cyan curves also nicely demonstrates how small the influence of the dissipative part of conductivity on the shape of transmission is, even for samples with a moderately high mobility around $10^6$ cm$^{2}$/Vs.
	
	The comparison presented in Fig.~\ref{FigS8} shows that despite overall good agreement between theory and experiment, the deviations between the low-$B$ decay of transmittance oscillations and MIRO are present, and can hardly be attributed solely to an incorrect choice of the fitting parameters. It would be interesting to check whether this inconsistency remains for other samples and conditions and to clarify the physical origin of the discrepancy.

	\newpage
	
\section{Conditions for observation of quantum transmittance oscillations}\label{Conditions}
	
Below we discuss and illustrate the conditions relevant for observation of the quantum transmittance oscillations in high-mobility 2DES, characterized by a large parameter $\omega\tau\equiv \mu B_\text{CR}\gg 1$ with $\tau$ being the transport scattering time. 
For simplicity, here we assume the ideal case of $K_\text{sw}=1$, 
and first consider the simplest case of constructive Fabry-P\'{e}rot interference, which is realized for integer $k d/\pi$ leading to $s_1=s_2=\pm 1$, so that the substrate becomes transparent for normally incident radiation at the corresponding frequency $\omega/2\pi$. The magnetotransmittance $|t_+|^2$, given by Eq.~(1) of the main text, is then fully determined by the dynamic conductivity $\sigma_+$ of 2DES,
\begin{equation}
    \label{Tsimple}
   |t_+|^2=\dfrac{1}{|1+\sigma_+ Z_0/2|^2}.
\end{equation}
In turn, the dynamic conductivity can be represented  as a series expansion,
\begin{equation}\label{series}
 \sigma_+= \dfrac{n e^2}{m\tilde{\omega}}\left(i+\dfrac{R_1}{\tilde{\omega}\tau}+\dfrac{i R_2}{(\tilde{\omega}\tau)^2}+\ldots\right) , \qquad  \tilde{\omega}\equiv\omega-\omega_c\gg 1/\tau.
\end{equation}
In the region of magnetic fields relevant for the transmittance oscillations the expansion parameter $\tilde{\omega}\tau\gg 1$, and the overall shape of transmission $|t_+|^2$ is dominated by the classical dynamical response of a disspationless 2DES as given by the first imaginary term in Eq.~(\ref{series}). The leading quantum corrections due to Landau quantization appear in the next-leading dissipative term given by \cite{dmitriev:2003}
\begin{equation}\label{R1}
 R_1= \int \text{d}\varepsilon \dfrac{f(\varepsilon) - f(\varepsilon+\hbar\omega)}{\omega} \tilde{\nu}(\varepsilon)\tilde{\nu}(\varepsilon+\omega)=1-4\delta\dfrac{X_T}{\sinh X_T}\dfrac{\omega_c}{2\pi\omega}\sin\dfrac{2\pi\omega}{\omega_c}\cos\dfrac{2\pi\varepsilon_\text{F}}{\hbar\omega_c}+2\delta^2\cos\dfrac{2\pi\omega}{\omega_c}.
\end{equation}
The integral over kinetic energies $\varepsilon$ above represents the statistical average of contributions of all possible impurity scattering processes involving emission or absorption of a photon with energy $\hbar\omega$. Function $f(\varepsilon)$ represents the thermal Fermi-Dirac distribution of electrons around the chemical potential $\varepsilon_\text{F}$. The probability of such emission and absorption processes is proportional to the product of initial and final density of states (DoS), which results in quantum oscillations reflecting the energy modulation of the normalized DoS, $\tilde{\nu}(\varepsilon)\equiv\nu(\varepsilon)/\nu_0\simeq 1 - 2\delta \cos(2\pi\varepsilon/\hbar\omega_\text{c})$. Here $\nu_0$ denotes the constant density of states at $B=0$, and the Dingle factor $\delta=\exp(-\pi/\mu_\text{q} |B|)$ is assumed to be small, reflecting the onset of Landau quantization corresponding to the limit of strongly overlapping Landau levels. 
The straightforward calculation then yields $R_1$ as a sum of the classical Drude value $R_1=1$, the dynamical analog  of the Shubnikov -- de Haas oscillations which exponentially decays at high $T$ where $X_\text{T} = 2 \pi^2 k_\text{B}T/\hbar \omega_\text{c}\gg 1$, and the second-order term $\propto\delta^2$ that survives at high $T$ similar to MIRO and overcomes the first-order term at $|B|<B_\text{CR}$ in conditions of our experiments. 
	\begin{figure}
		\includegraphics[width=1\columnwidth]{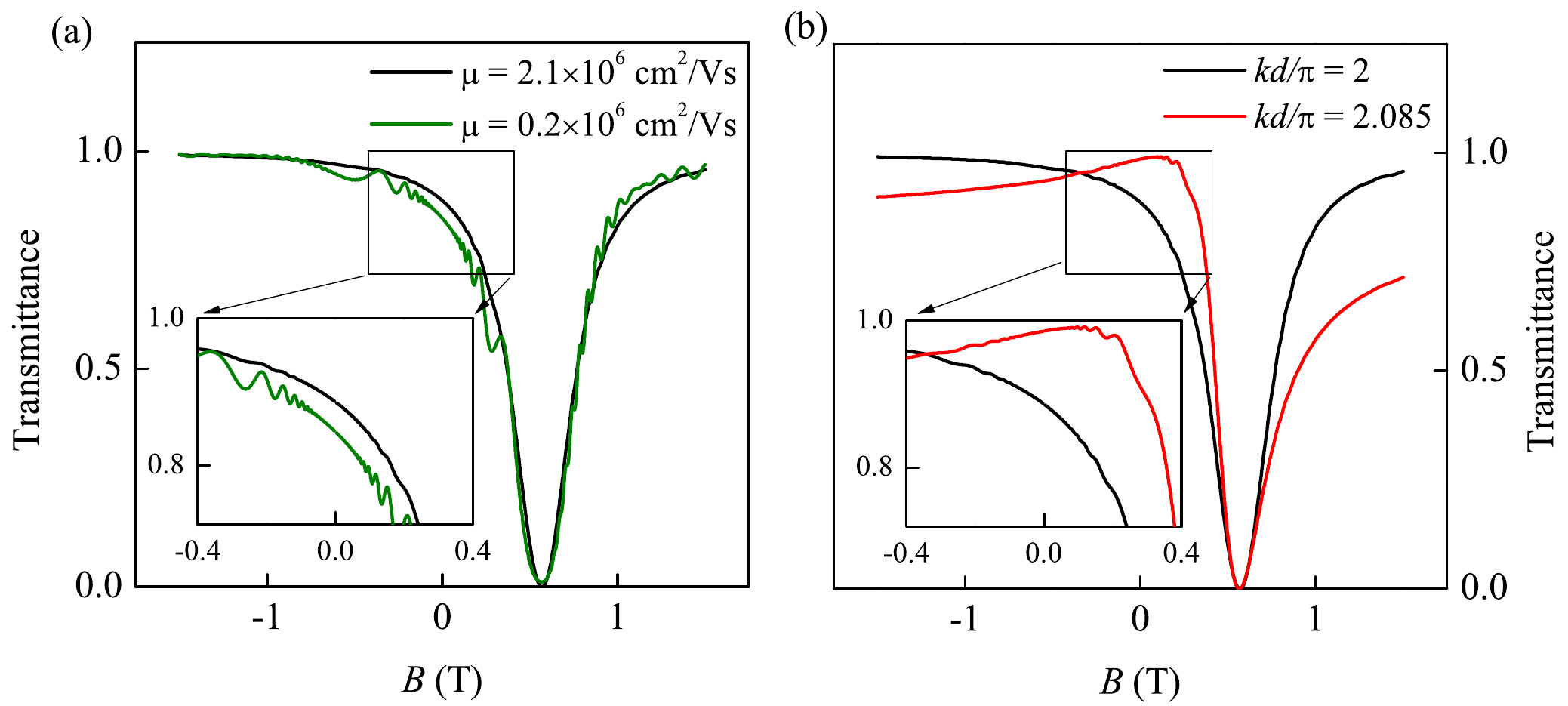}
		\caption{
			Calculated magnetic field dependences of transmittance $|t_+|^2(B)$ that demonstrate the influence of mobility (a) and interference phase (b) on the visibility of quantum oscillations.	The red line in panel (b) reproduces the fit to measured transmittance signal in Fig.~1 of the main text. The black line in both panels is calculated using the same parameters, apart from the interference phase which is set to integer $k d/\pi$ to match the condition of constructive Fabry-P\'{e}rot interference. The asymmetric shape of the transmittance helps to resolve quantum oscillations on top of an almost flat shoulder around $B=0$. The green line in panel (a) demonstrates the emergence of more pronounced quantum oscillations in a system with approximately ten times lower mobility.
		} \label{FigS9}
	\end{figure}

While quantum effects in Eq.~(\ref{R1}) become well pronounced at $\mu_\text{q} |B|\sim 1$ (we estimate $2\delta^2\sim 0.5$ at $|B|=0.2$\,T for data shown in Fig.~1 and modelled below), their detection in the transmittance remains a challenging task owing to the large value of $\omega\tau$. 
This is illustrated in Fig.~\ref{FigS9}\,(a), where the magnetotransmission calculated according to Eq.~(\ref{Tsimple}) -- (\ref{R1}) is shown for $\mu = 2.1 \times 10^6\,$cm$^2$/Vs (black line) and $\mu = 0.2 \times 10^6\,$cm$^2$/Vs  (green line). 
The other parameters including the frequency $f = 222\,$GHz, density $n = 6.6 \times 10^{11}\,$cm$^{-2}$, temperature $T=1.9$~K, and quantum mobility $\mu_\text{q} = 0.2 \times 10^6\,$cm$^2$/Vs are fixed at values corresponding to those in Fig.~1 of the main text. 
While the green, low-mobility curve ($\omega\tau\simeq12$) shows prominent quantum oscillations (dominated by the last term in Eq.~(\ref{R1}) at $|B|<B_\text{CR}$ and by the preceding Shubnikov -- de Haas term at $|B|>B_\text{CR}$), in the black curve, corresponding to the parameters of our sample ($\omega\tau\simeq120$), these oscillations become strongly suppressed and are barely visible. Large typical values of $\omega\tau$ explain why no magnetooscillations in transmittance were reported previously for ulta-high-mobility samples, such as one with $\mu = 18 \times 10^6\,$cm$^2$/Vs ($\omega\tau\sim 1000$) studied in Ref.~\onlinecite{smet:2005}. 

The observation of small oscillations on top of smooth but rapidly changing background in samples with moderately high mobility, see black curve in Fig.~\ref{FigS9}\,(a), is still a technically nontrivial task. We have found that an opportunity to control the phase of Fabry-P\'{e}rot interference by a proper choice of the radiation frequency helps to overcome this difficulty. The influence of the interference phase is illustrated in the right panel of Fig.~\ref{FigS9}. Here the black curve, calculated using Eq.~(\ref{Tsimple}) for the case of constructive interference, is the same as in the left panel. 
The red curve reproduces the modelled transmission in Fig.~1 of the main text, which is calculated using the general expression (1) of the main text with $K_\text{sw}=1$. 
Here, the interference parameter $k d= 2.085\,\pi$ is obtained for the actual sample thickness $d = 406\,\mu$m and for the chosen radiation frequency $f = 222\,$GHz. 
This choice leads to appearance of an almost flat region in the background transmission signal that improves the visibility of magnetooscillations.

	\newpage
	
	\bibliography{newbib}